\documentstyle[aps,prl,epsfig,multicol]{revtex}
\input epsf

\begin{document}
\draft
\title{Formation of a stable decagonal 
quasicrystalline Al-Pd-Mn surface layer}
\author{D. Naumovi\'{c},$^{1}$ P. Aebi,$^{1}$ L. Schlapbach,$^{1}$ 
C. Beeli,$^{2}$ K. Kunze,$^{3}$ T. A. Lograsso,$^{4}$ D. W. Delaney $^{4}$ }
\address{$^{1}$Institut de Physique, Universit\'e de Fribourg, P\'erolles, CH-1700 
Fribourg, Switzerland}
\address{$^{2}$Laboratory of Solid State Physics, ETHZ, CH-8093 Z\"urich, Switzerland}
\address{$^{3}$Institute of Geology, ETHZ, CH-8092 Z\"urich, Switzerland}
\address{$^{4}$Ames Laboratory, Iowa State University, Ames, Iowa 50011, USA}

\maketitle
\begin{abstract}
We report the $in$ $situ$ formation of an ordered equilibrium decagonal Al-Pd-Mn 
quasicrystal overlayer on the 5-fold symmetric surface of an icosahedral 
Al-Pd-Mn monograin. The decagonal structure of the epilayer is 
evidenced by x-ray photoelectron diffraction, low-energy electron 
diffraction and electron backscatter diffraction. This overlayer is 
also characterized by a reduced density of states near the Fermi 
edge as expected for quasicrystals. This is the first time that a 
millimeter-size surface of the stable decagonal Al-Pd-Mn is obtained, 
studied and compared to its icosahedral counterpart.
\end{abstract} 

\begin{multicols}{2}
\narrowtext	
%
%
Quasicrystals are exceptional in many aspects.  They emerged from the 
rather complex Al-Mn phase diagram at the beginning of the 80's 
\cite{sche}.  Their structure is infringing the classic rules of 
crystallography: they are non periodic, but long-range ordered, and 
they include forbidden symmetry axes such as five-fold (5$f$), 8$f$, 
10$f$ or 12$f$, depending on the alloy\cite{janot}.  Furthermore, the 
electronic structure of quasicrystals is quite unexpected.  The 
electrical resistivity is a thousand times higher than their 
constituents, metals, and inversely proportional to 
temperature\cite{poon}.  Despite the lack of periodicity a band-like 
behavior is observed in the electronic structure \cite{Wu,Rotenberg}.  
Quasicrystals also exhibit properties which are directly attractive 
for applications, such as low friction, low adhesion or increased 
hardness\cite{dubois}.

Even today, producing quasicrystals, and especially monograin samples 
of stable phases, remains a challenge.  In that sense, one of the most 
fruitful systems is Al-Pd-Mn.  In 1990, Tsai et al.  discovered a 
first stable icosahedral ($i$) quasicrystal ($i$-Al$_{\rm 70}$Pd$_{\rm 
20}$Mn$_{\rm 10}$) in the ternary Al-Pd-Mn system\cite{Tsai1}.  It is 
possible to produce high-quality large monograins of this alloy.  This 
facilitates experiments, such as neutron or x-ray diffraction, or 
surface experiments, to be performed on quasicrystals.  A $stable$, 
$decagonal$ ($d$) Al-Pd-Mn quasicrystal was found by Beeli, Nissen and 
Robadey\cite{beeli}.  It exists only in a very narrow area in the 
phase diagram\cite{Godecke}.  Therefore, it is difficult to obtain a 
single-phase specimen with high quality of the decagonal structure.  
Interestingly, a decagonal quasicrystal with high structural quality 
can be obtained by annealing rapidly quenched tapes of an Al$_{\rm 
69.8}$Pd$_{\rm 12.1}$Mn$_{\rm 18.1}$ alloy, which contains a 
metastable icosahedral phase, but has the equilibrium composition of 
the decagonal phase\cite{beeli}.  Thus the icosahedral long-range 
order is coherently transformed into decagonal long-range order.  
Furthermore, single-grain samples can only be produced by 
non-equilibrium processes via metastable reactions, and, generally, 
only by chance single grains with diameters of 0.1 mm are produced.  
Nevertheless, Al-Pd-Mn has the only known phase diagram containing two 
stable quasicrystalline phases, the icosahedral and the decagonal.

Until now, many surface experiments have been reported on monograin 
$i$-Al-Pd-Mn.  Clean surfaces are prepared either by 
fracturing\cite{Ebert,Neuhold,Fournee} or by ion sputtering, and by annealing.  
Composition changes are induced by preferential sputtering of the 
lightest elements, or by thermal diffusion.  For annealing 
temperatures lower than 400${\rm^{o}}$C subsequent or simultaneous to 
sputtering, a crystalline phase with cubic domains (Al$_{\rm 
55}$Pd$_{\rm 40}$Mn$_{\rm 5}$) 
\cite{Shen,NauICQ6,BollPRL,NauRapid,NauICQ7,Schmithusen} or a 
$metastable$ decagonal quasicrystalline (Al$_{\rm 22}$Pd$_{\rm 
56}$Mn$_{\rm 22}$) phase\cite{BolldAPM} was observed, respectively.  
Further annealing of these surfaces at temperatures between approximately 
450${\rm^{o}}$ and 650${\rm^{o}}$C restores a bulk-terminated 
icosahedral surface and a bulklike composition.  Such surfaces are 
characterized by a reduced density of states close to the Fermi level 
($E_{F}$), in contrast to usual metallic surfaces 
\cite{Wu,Fournee,NauRapid,NauICQ7,Berger}. Annealing at higher 
temperatures produces either Pd or Mn enrichment of the surface, 
probably depending on the initial bulk composition.  An irreversible 
precipitation of secondary phases is even reported \cite{Thiel}.  The 
Pd-rich surfaces Al$_{\rm >66}$Pd$_{\rm 32}$Mn$_{\rm <2}$ 
\cite{NauICQ6,NauSS} or Al$_{\rm 50}$Pd$_{\rm 49}$Mn$_{\rm 
1}$\cite{Schmithusen} were interpreted as cubic with domains and the 
Mn-rich Al$_{\rm 75}$Pd$_{\rm 6}$Mn$_{\rm 19}$ surface\cite{Ledieu} as 
corresponding to the crystalline orthorhombic Al$_{\rm 3}$Mn.  Both, 
low- and high-temperature crystalline phases are epitactically grown 
on the substrate and rather ordered.

In this work, we produce for the first time a large surface of the 
$stable$ decagonal quasicrystalline $d$-Al-Pd-Mn.  The 10$f$-symmetric 
surface of the stable $d$-Al-Pd-Mn is epitactically grown as a 
single-domain overlayer on the 5$f$-symmetric $i$-Al-Pd-Mn, with the 
10$f$ axis of the overlayer parallel to the 5$f$ axis of the 
substrate.  The Mn-rich decagonal overlayer is produced $in$ $situ$ 
by sputtering and annealing the 5$f$ $i$-Al-Pd-Mn at about 
650${\rm^{o}}$C. The results on geometrical (Fig. 1) and electronic 
structure (Fig. 2), obtained with low-energy electron diffraction 
(LEED), x-ray photoelectron diffraction (XPD), electron backscatter 
diffraction (EBSD) and ultraviolet photoemission (UPS), clearly 
demonstrate that both the icosahedral surface and the epitactically 
grown decagonal overlayer can be identified with the corresponding 
stable bulk phases.

%
%
The photoemission experiments were performed in a VG ESCALAB Mk~II 
spectrometer with a base pressure in the $10^{\mathrm{-11}}$~mbar 
range.  The sample stage is modified for motorized sequential 
angle-scanning data acquisition over the full solid angle 
\cite{osterw91e}.  MgK$_{\mathrm{\alpha}}$ radiation ($h\nu$=1253.6 
eV) was used for x-ray photoelectron spectroscopy (XPS) in order to 
check the cleanliness of the sample and to determine the 
composition\cite{xps} within the probing depth (mean free path of 
photoelectrons $\sim 20$ \AA).  XPD served to study the geometrical 
structure of the surface \cite{fadley90}.  In angle-scanned XPD, the 
intensity variations of photoemitted core-level electrons are recorded 
as a function of emission angle and stereographically projected in a 
greyscaled map.  For kinetic energies $>$500 eV, photoelectrons 
leaving the emitter atom are strongly focused in the forward direction 
by neighboring atom potentials \cite{fadley90}.  So, the intensity is 
enhanced along densely packed atomic rows or planes.  Thus, XPD 
evidences the average local order around a selected chemical species.  
UPS measurements are performed with monochromatized He~I$\alpha$ 
radiation (21.2~eV)\cite{pillo98e}.  The energy resolution of the 
analyzer for the UPS measurements was set to 30~meV. All measurements 
are performed at room temperature.

The symmetry and orientation of substrate and layer were verified $ex$ 
$situ$ by EBSD \cite{EBSD1}.  EBSD patterns were recorded using a 
standard SEM CamScan CS44LB, equipped with a 50 mm diameter area 
detector (scintillator and high sensitivity TV camera) placed parallel 
to the incident beam.  Diffraction patterns are formed from the 
spatial intensity distribution of backscattered electrons, which are 
induced by a focused stationary electron beam (15 kV, 3nA).  The 
sample surface is tilted 70${\rm^{o}}$ against the beam towards the 
detector.  The spatial resolution is in the range of some tenth of a 
micrometer and anisotropic due to the 70${\rm^{o}}$ tilt.  Bands of 
higher intensity arise between pairs of pseudo-Kikuchi lines, the 
center of which marks the projection of diffracting symmetry planes 
onto the detector screen.  These bands are detected in digitized EBSD 
patterns by image processing using a Hough transform \cite{EBSD2}.  
The software OIM2.0 (TSL Inc.)  was modified by us incorporating 
quasicrystallography, so that the model pattern could be recalculated 
which fits best with the detected diffraction bands for EBSD patterns 
of any orientation and symmetry.

The $i$-Al-Pd-Mn quasicrystal ingot was grown using the Bridgman 
method \cite{delaney}. Its bulk composition was 
determined to be Al$_{\rm 71.6}$Pd$_{\rm 19.6}$Mn$_{\rm 8.6}$.  The 
sample was an 1.3 mm thick disk with a diameter of 8 mm.  It was 
oriented perpendicular to a 5$f$ symmetry axis within 0.25${\rm^{o}}$ 
and polished with diamond paste and colloidal silica.
\begin{figure}[b]
\centerline{\epsfig{file=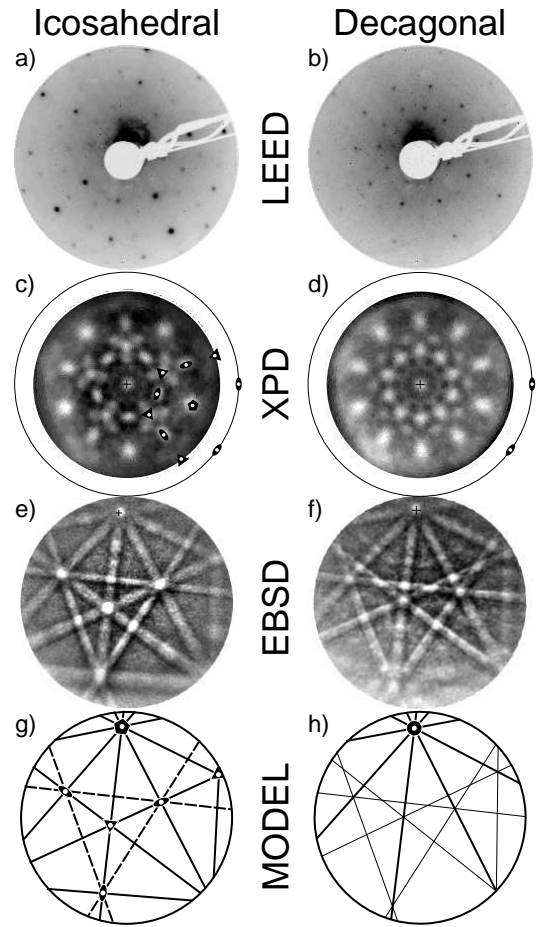,width=7cm}}
	\vspace{0.3cm} 
	\caption{LEED patterns taken at 60 eV (high 
	intensity in black) from a monograin $i$-Al-Pd-Mn cut 
	perpendicularly to a 5$f$-symmetry axis: (a) the bulk-terminated 
	icosahedral quasicrystalline {\em i}-Al$_{\rm 68}$Pd$_{\rm 
	26}$Mn$_{\rm 6}$ surface obtained after sputtering and annealing 
	at 550${\rm^{o}}$C; (b) the stable decagonal quasicrystalline 
	$d$-Al$_{\rm 76}$Pd$_{\rm 11}$Mn$_{\rm 13}$ overlayer obtained 
	after sputtering and annealing at 650${\rm^{o}}$C. XPD patterns of 
	Pd $3d_{5/2}$ (E$_{\rm{kin}}$ = 915 eV, high intensity in white) 
	with indicated normal emission (+), grazing emission (outer 
	circle), 2$f$- (ellipse), 3$f$- (triangle), 5$f$ (pentagon) 
	-symmetry axes, taken from, (c), the icosahedral surface described 
	in (a); and, (d), the decagonal surface described in (b).  EBSD patterns 
	(high intensity in white; sample normal (+)) taken from, (e), the 
	icosahedral surface described in (a); and, (f), the decagonal surface 
	described in (b).  Projections generated by a pattern recognition 
	procedure, with indicated 2$f$-, 3$f$-, 5$f$- (as above), 10$f$ 
	(decagon) -symmetry axes, and 2$f$- (solid thick line), 5$f$ 
	(dashed thick line)-symmetry planes, (g) identifying (e) as 
	quasicrystalline icosahedral; (h) identifying (f) as 
	quasicrystalline decagonal, with additional 1st order rhombohedral planes 
	(thin solid lines).}
	\label{fsm}
\end{figure}
%
%
For the preparation of the icosahedral surface, sputtering and 
annealing to 550${\rm^{o}}$C resulted in XPS compositions of Al$_{\rm 
53}$Pd$_{\rm 41}$Mn$_{\rm 6}$ and Al$_{\rm 68}$Pd$_{\rm 26}$Mn$_{\rm 
6}$, respectively.  However, for annealing temperatures of 
650${\rm^{o}}$C, the surface becomes decagonal with a modified 
composition of Al$_{\rm 76}$Pd$_{\rm 11}$Mn$_{\rm 13}$.  In contrast 
to a metastable decagonal phase Al$_{\rm 22}$Pd$_{\rm 56}$Mn$_{\rm 
22}$ (an icosahedral surface is recovered after annealing above 
400${\rm^{o}}$C, Ref.  \cite{BolldAPM}), the composition is, in our 
case, compatible with the stable decagonal quasicrystal composition 
(Al$_{\rm 69.8}$Pd$_{\rm 12.1}$Mn$_{\rm 18.1}$)\cite{beeli}.  In 
addition, the decagonal overlayer, once created, remains stable up to 
700${\rm^{o}}$C. After 15 to 30 min sputtering the decagonal surface, 
the composition is Al$_{\rm 66}$Pd$_{\rm 15}$Mn$_{\rm 19}$ and a 5 min 
annealing at 650${\rm^{o}}$C is sufficient to recover the decagonal 
surface.  In contrast, typically 5h of sputtering followed by 
annealing at 550${\rm^{o}}$C are necessary to recover the icosahedral 
bulk-terminated surface.  Therefore, transformation from the decagonal 
back to the icosahedral surface is also possible.

Figure 1 displays the experimental data comparing the geometrical 
structure of the icosahedral bulk-terminated surface, and of the 
stable decagonal overlayer, measured with LEED, XPD and EBSD. The 
bulk-terminated icosahedral surface is characterised by a 5$f$ LEED 
pattern (Fig.  1(a)), while the LEED picture of the decagonal 
overlayer (Fig.  1(b)) is 10$f$ symmetric.  This latter displays four 
rings of ten fine spots.  The relationship between the radii of these 
four rings is proportional to the golden mean $\tau$, confirming a 
quasicrystalline ordering of the overlayer.  LEED spots are visible 
over a wide range of energies (12-140 eV).  This indicates the 
long-range lateral ordering of the overlayer 10$f$ surface, compatible 
with a stable decagonal quasicrystalline phase.  Note, that this 
pattern drastically differs from the one taken from the low- and 
high-temperature pseudo-10$f$ surface with crystalline domains 
\cite{Shen,NauICQ6,Schmithusen,NauSS}.  Decagonal LEED patterns have 
been obtained from the $d$-Al-Ni-Co \cite{Gierer}.

The Pd $3d_{5/2}$ XPD pattern of the bulk-terminated 5$f$ surface is 
shown in Fig.  1(c).  The main high-intensity spots can be easily 
identified with the 2$f$-, 3$f$- and 5$f$-symmetry axes of the 
icosahedral group projected stereographically 
\cite{NauICQ6,NauICQ7,NauAmes,NauSS}, with a 5$f$-symmetry axis in the 
center (normal emission).  These experimental patterns can be 
reproduced by single-scattering cluster calculations using clusters as 
inferred from structural models of the $i$-Al-Pd-Mn phase, confirming 
that the near surface region is quasicrystalline\cite{NauAmes}.  So we 
conclude that the average local environment around the Pd emitter 
atoms is icosahedral (idem for Al and Mn, not shown\cite{NauAmes}).  
In contrast, Fig.  1(d) displays the XPD pattern of the Mn-rich stable 
decagonal overlayer, with an evident overall 10$f$ symmetry and a 
central 10$f$-symmetry axis.  Here, three rings of equivalent intense 
spots replaces the alternation of spots identified as 2$f$-, 3$f$- and 
5$f$-symmetry axes in Fig.  1(c).  This clearly indicates an average 
short-range decagonal ordering of the overlayer.  Strong 
epitactic-like relationships exist between the icosahedral bulk and 
the decagonal overlayer.  The decagonal XPD pattern resembles the XPD 
\cite{ShimodaSS} and secondary electron imaging (SEI) \cite {Zurkirch} 
patterns taken from $d$-Al-Ni-Co.  Note that the SEI pattern of the 
decagonal metastable Al$_{\rm 22}$Pd$_{\rm 56}$Mn$_{\rm 22}$ overlayer 
resembles the sputtered $d$-Al-Ni-Co surface \cite{BolldAPM}.  The XPD 
patterns of Pd-rich pseudo-10$f$ surface \cite{NauICQ6,NauSS} and of 
sputtered $d$-Al-Ni-Co surface\cite{ShimodaSS} were both interpreted 
as crystalline with bcc domains.

Figures 1(e-f) show the EBSD results obtained $ex$ $situ$ from the 
icosahedral and decagonal phase created on 5$f$ $i$-Al-Pd-Mn sample.  
The EBSD pattern presented in Fig.  1 (e) is identified as 
corresponding to icosahedral phase (Fig.  1(g)) with a 5$f$-symmetry 
axis coinciding with the sample normal (marked by +).  Further 2$f$- 
and 3$f$-symmetry axes are situated at crossings of 2$f$- and 
5$f$-symmetric planes as indicated in Fig.  1 (g).  These latter 
symmetry axes, as well as 5$f$-symmetric planes, do not exist in the 
pattern (Fig.  1(f)) identified as decagonal (Fig.  1(h)).  Here, a 
10$f$-symmetry axis coincides with the sample normal, and only the 
radial 2$f$-symmetric Kikuchi bands crossing the 10$f$ axis persist.  
As EBSD arises from 200 to 500 \AA$ $ depth, the thickness of the 
stable decagonal phase is at least of that order of magnitude.  Once 
the decagonal overlayer formed, the appearance of the icosahedral 
pattern (alone or superimposed to the decagonal pattern with lower 
contrast) indicates that some patches ($<$1$\%$) of the surface are 
covered with a thinner decagonal overlayer or remain uncovered.  In 
addition, this evidences the strong structural and orientational 
relationship between the icosahedral bulk and the decagonal overlayer.
\begin{figure}[b]
\centerline{\epsfig{file=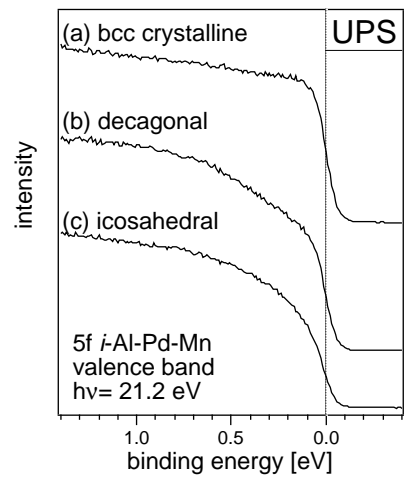,width=6cm}}
	\vspace{0.3cm} 
	\caption{Room-temperature valence-band spectra 
	displaying the near $E_{F}$ region, taken with monochromatized 
	He-I radiation (h$\nu$ = 21.2 eV), (a) of the crystalline surface with bcc 
	domains (Al$_{\rm 53}$Pd$_{\rm 41}$Mn$_{\rm 6}$; 
	sputtered), (b) of the Mn-rich stable decagonal quasicrystalline 
	overlayer (Al$_{\rm 76}$Pd$_{\rm 11}$Mn$_{\rm 13}$; sputtered and 
	annealed at 650${\rm^{o}}$C) and (c) of the bulk-terminated 
	icosahedral quasicrystalline surface (Al$_{\rm 68}$Pd$_{\rm 
	26}$Mn$_{\rm 6}$; sputtered and annealed at 550${\rm^{o}}$C).}
	\label{vb}
\end{figure}
Figure 2 displays valence-band spectra taken at room temperature of 
the icosahedral bulk-terminated surface, of the stable decagonal 
overlayer and of a crystalline phase, obtained by sputtering the 
icosahedral surface.  This latter 5$f$ surface consists of five 
bcc(113) domains rotated by 72${\rm^{o}}$ with respect to each other as seen 
with XPD\cite{NauICQ7}.  The density of states remains high and linear 
over the complete range of energy and the Fermi edge is sharp, as 
expected from a metallic surface (Fig.  2a).  In Fig.  2(c), the 
bulk-terminated icosahedral surface exhibits a completely different 
behaviour.  A distinct decrease of the density of states near $E_{F}$, 
interpreted as the opening of a pseudogap, is observed, as expected 
from a quasicrystalline surface 
\cite{Wu,NauRapid,NauICQ7,Berger,stadnikPRL}.  Finally, the density of 
states of the stable decagonal overlayer (Fig.  2 (b)) is lowered 
close to $E_{F}$, as for the icosahedral quasicrystalline surface.  
However, the shape of the curve is slightly different, with a steeper 
Fermi cutoff.  This is probably related to the fact that the decagonal 
surface is periodic in one dimension (along the surface normal) and 
that the quantity of Mn (Mn 3$d$ states determine the spectral weight 
close to $E_{F}$ \cite{stadnikPRL}) is doubled in the decagonal phase 
compared to the icosahedral and crystalline bcc phases.

An important point to be discussed is how to ensure the 
quasicrystallinity of the surface. In previous work on 
$i$-Al-Pd-Mn\cite{NauRapid,NauICQ7}, we demonstrated that a 
combination of geometrical and electronic structure techniques can 
efficiently characterize the quasicrystal surfaces.  First, the ordering and 
the symmetry within differently prepared surfaces were probed with 
XPD and LEED.  But note, that a 5$f$\cite{NauICQ7} or 
10$f$\cite{NauICQ6,NauSS} symmetry can also be due to a combination of 
domains and not only to the intrinsic quasicrystal symmetry.  Second, 
a distinct suppression of spectral weight of the density of states close 
to $E_{F}$ was observed on quasicrystalline terminated surfaces 
by UPS.  Here, one has to be aware that a suppressed 
density of states is also reported on a crystalline approximant of 
quasicrystal\cite{Fournee}. Whereas, even a slight 
disordering of the quasicrystal surface, induced by a one-minute ion 
sputtering, produces a sharp Fermi edge cutoff, characteristic of a metallic
surface, with no apparent composition 
changes\cite{Berger}.  So, a careful evaluation of the 
geometrical and electronic structure data, together with the 
composition, is essential to identify quasicrystallinity or,
at least, to exclude crystallinity of the surface. 

%
%
In summary, this work presents for the first time the formation of an 
extended and ordered stable decagonal quasicrystalline overlayer on 
5$f$ $i$-Al-Pd-Mn by sputtering and annealing at 650${\rm^{o}}$C. This 
overlayer was characterized, at different scales, with three structure 
techniques, XPD, LEED and EBSD. We clearly identified the icosahedral 
and decagonal symmetries of the surfaces and evidenced the 
epitactic-like relationships between the decagonal overlayer and the 
icosahedral bulk.  Furthermore the decagonal overlayer exhibited a 
suppressed density of states near $E_{F}$ as seen with UPS , as 
expected for quasicrystals.  The production of the two 
quasicrystalline, decagonal and icosahedral, alloys of the same 
elemental Al-Pd-Mn family gives the unique opportunity to study and 
compare the properties of these two phases.

With grateful thanks to M. Boudard, Ph.  Ebert, T. Janssen, P. Thiel 
for fruitful discussions; to the members of the XPD team and the 
technical staff for help; and to the Swiss National Science Foundation 
for financial support.

 
\end{multicols}

\begin{thebibliography}{10}
\bibitem{sche}
D. Shechtman $et$ $al.$, Phys.  Rev.  Lett.  \textbf{53}, 1951 
(1984).

\bibitem{janot}
C. Janot, Quasicrystals: A Primer, 2nd ed.  (Oxford University Press, 
Oxford, 1994).

\bibitem{poon}
S.J. Poon, Adv.  Phys.  \textbf{41}, 303 (1992).

\bibitem{Wu}
X. Wu $et$ $al.$, Phys.  Rev.  Lett.  \textbf{75}, 4540 (1995).

\bibitem{Rotenberg} E. Rotenberg $et$ $al.$, Nature \textbf{406}, 602 
(2000).

\bibitem{dubois}
J.-M. Dubois, Phys.  Scr.  \textbf{T49}, 17 (1993).

 \bibitem{Tsai1} A. P. Tsai $et$ $al.$, Philos.  Mag.  Lett.  
 \textbf{61}, 9 (1990).

 \bibitem{beeli} C. Beeli, H.-U. Nissen, and J. Robadey, Philos.  Mag.  
 Lett.  \textbf{63}, 87 (1991); C. Beeli $et$ $al.$, in Proceedings of the 5th 
 International Conference on Quasicrystals, Avignon, 1994, edited by 
 C. Janot and R. Mosseri (World Scientific, Singapore, 1995) p.  680.
 
  \bibitem{Godecke} T. G\"{o}decke and R. L\"{u}ck, Z. 
 Metallk.\textbf{86}, 109 (1995).

 \bibitem{Ebert}Ph.  Ebert $et$ $al.$, Phys.  Rev.  Lett.  
 \textbf{77}, 3827 (1996); Phys.  Rev.  B \textbf{57}, 2821 (1998); 
 ibid.  \textbf{60}, 874 (1999).
 
\bibitem{Neuhold}
G. Neuhold $et$ $al.$, Phys.  Rev.  B.  \textbf{58}, 734 (1998).
 
 \bibitem{Fournee}V. Fourn\'{e}e $et$ $al.$, Phys.  Rev.  B 
 \textbf{62}, 14049 (2000).

 \bibitem{Shen} Z. Shen $et$ $al.$, Phys.  Rev.  Lett.  
 \textbf{78},1050 (1997); Phys.  Rev.  B \textbf{58}, 9961 (1998); 
 Surf.Sci.  \textbf{450}, 1 (2000).

 \bibitem{NauICQ6}D. Naumovi\'c $et$ $al.$, in Proceedings of the 6th 
 International Conference on Quasicrystals, Tokyo, 1997, edited by S. 
 Takeuchi and T. Fujiwara (World Scientific, Singapore,1998) p.  749.

 \bibitem{BollPRL} B. Bolliger $et$ $al.$, Phys.  Rev.  Lett.  
 \textbf{80}, 5369 (1998).

\bibitem{NauRapid} D. Naumovi\'{c} $et$ $al.$, Phys.  Rev.  B 
\textbf{60}, R16330 (1999).

\bibitem{NauICQ7} D. Naumovi\'{c} $et$ $al.$, Mat.  Sci.  and Eng.  A. 
\textbf{294-296}, 882 (2000).

\bibitem{Schmithusen} F. Schmith\"{u}sen $et$ $al.$, Surf.  Sci.  
\textbf{444}, 113 (2000).


 \bibitem{BolldAPM} B. Bolliger $et$ $al.$, Phys.  Rev.  Lett.  
 \textbf{82}, 763 (1999).


 \bibitem{Berger} T. Schaub $et$ $al.$, Eur.  Phys.  J. B 
 \textbf{20},183 (2001).
 

\bibitem{Thiel} P. A. Thiel, in Physical Properties of Quasicrystals 
edited by Z.M. Stadnik (Springer-Verlag, Berlin, 1999), p.  327.

\bibitem{NauSS} D. Naumovi\'{c} $et$ $al.$, Surf.  Sci.  
\textbf{433-435}, 302 (1999).


 \bibitem{Ledieu} J. Ledieu $et$ $al.$, Mat.  Sci.  and Eng.  A 
 \textbf{294-296}, 871 (2000).
 
 \bibitem{osterw91e} J. Osterwalder $et$ $al.$, Phys.  Rev.  B {\bf 
 44}, 13764 (1991).
 
  \bibitem{xps}Note that XPS concentrations cannot directly be compared 
 with a true stoichiometry because it is only based on intensity ratios 
 including core-level cross sections; see \onlinecite{NauICQ6,NauSS}.
 
 \bibitem{fadley90}See, e.g., C.S. Fadley, Surf.  Sci.  
Rep.  {\bf 19}, 231 (1993).

 \bibitem{pillo98e} Th.  Pillo $et$ $al.$, J. Electr.  Spectrosc.  
 Relat.  Phenom.  {\bf 97}, 243 (1998).
 
 \bibitem{EBSD1} J. A. Venables, and C. J. Harland, Phil. Mag. 
\textbf{27}, 1193 (1973).

\bibitem{EBSD2} B. L. Adams, S. I. Wright, and K. Kunze, Met. Trans. 
\textbf{24A}, 819 (1993); K. Kunze $et$ $al.$, Text. Microstruct. 
\textbf{20}, 41 (1993). 
 
 \bibitem{delaney} D. W. Delaney, T. E. Bloomer, and T. A. Lograsso, 
 in New Horizons in Quasicrystals (Ref. 
 \onlinecite{NauAmes}) p.45.
 
 \bibitem{Gierer} M. Gierer $et$ $al.$, Surf.  Sci.  \textbf{463}, L654 
(2000). 
 
 \bibitem{NauAmes} D.  Naumovi\'c $et$ $al.$, in New Horizons in Quasicrystals, 
 edited by A.I.Goldman $et$ $al.$ (World Scientific, Singapore, 1997) p.86 (For 
 greyscale Figs. 3 and 4, ask ``Dusanka.Naumovic@unifr.ch'').

\bibitem{ShimodaSS} M. Shimoda $et$ $al.$, Surf.  Sci.  
 \textbf{454-456}, 11 (2000).
 
 \bibitem{Zurkirch}M. Zurkirch, M. Erbudak, and A. R. Kortan, in 
 Proceedings of the 6th International Conference on Quasicrystals (Ref. 
 \onlinecite{NauICQ6}) p.  67; M. Zurkirch $et$ $al.$, Phys.  Rev.  B 
 \textbf{58}, 14113 (1998).

 \bibitem{stadnikPRL}
Z.M. Stadnik $et$ $al.$, Phys.  Rev.  Lett.  \textbf{77}, 1777 (1996).


\end{thebibliography}
\end{document}